\documentclass[namedreferences,hyperref,optionalrh]{spr-sola}
\usepackage{graphicx}        % For eps figures, newer & more powerfull
\usepackage{color}           % For color text: \color command
\usepackage{amsmath}
\renewcommand{\vec}[1]{{\mathbfit #1}}

\newcommand{\eq}[1]{(\ref{#1})}

\newcommand{\Tab}[1]{Table~\ref{#1}}
\newcommand{\Fig}[1]{Figure~\ref{#1}}
\newcommand{\Figs}[2]{Figures~\ref{#1} and \ref{#2}}

\newcommand{\mps}{m~s$^{-1}$}

\newcommand{\cmss}{cm$^2$~s$^{-1}$}
\chardef\us=`\_
\def\bl{Babcock--Leighton}
\def\mc{meridional circulation}

\begin{document}

\begin{frontmatter}
\title{Analyses of features of magnetic cycles at different amounts of dynamo supercriticality: Solar dynamo is about two times critical}
\author[addressref={aff1},corref,email={wavhal.sankets.phy20@itbhu.ac.in}]{\inits{S.}\fnm{Sanket}~\lnm{Wavhal}}%\sep
\author[addressref={aff1,aff2},corref,email={pawan.kumar.rs.phy18@itbhu.ac.in, pawankumar@iiap.res.in}]{\inits{P.}\fnm{Pawan}~\lnm{Kumar}\orcid{ 0009-0008-0946-0341}}%\sep
%\author[addressref=aff2,corref,email={pawankumar@iiap.res.in}]{\inits{P.}\fnm{Pawan }~\lnm{Kumar}}
\author[addressref=aff1,corref,email={karak.phy@itbhu.ac.in}]{\inits{B.B.}\fnm{Bidya Binay}~\lnm{Karak}\orcid{987-654-3210}}

\address[id=aff1]{Department of Physics, Indian Institute of Technology (BHU) Varanasi 221005, India}
\address[id=aff2]{Indian Institute of Astrophysics, Koramangala, Bengaluru, 560034, India}
%\address[id=aff3]{Third affiliation and address}

\runningauthor{Wavhal et al.}
\runningtitle{Supercriticality of solar dynamo}

\begin{abstract}
The growth of a large-scale magnetic field in the Sun and stars is usually possible when the dynamo number $(D)$ is above a critical value $D_c$. As the star ages, its rotation rate and thus $D$ decrease. Hence, the question is how far the solar dynamo is from the critical dynamo transition. 
To answer this question, we have performed a set of simulations using \bl\ type dynamo models at different values of dynamo supercriticality and analyzed various features of magnetic cycle. By 
%(i) 
comparing  the recovery rates of the dynamo from the Maunder minimum and statistics (numbers and durations) of the grand minima and maxima with that of observations and 
%(ii) modeling the amplitudes of the last five solar cycles using the observed polar field, 
we show that the solar dynamo is only about two times critical and thus not highly supercritical.
The observed correlation between the polar field proxy and the following cycle amplitudes and Gnevyshev-Ohl rule are also compatible with this conclusion.
\end{abstract}
\keywords{Solar dynamo; convection zone; magnetic field; solar cycle; sunspot}
\end{frontmatter}
%-------------------------------------------------

\section{Introduction}
\label{Introduction} 

Observations show that the Sun and Sun-like stars exhibit variable large-scale magnetic fields. 
This variation in the magnetic fields can be best seen in the form of irregular amplitude, grand minima and maxima, Gnevyshev-Ohl rule, Gnevyshev peaks
\citep{ Wright16, Shah18, KMB18, garg19, Uso23, Karak23}. 
For the Sun, these characteristics can be observed in the last 400 years sunspot data and its proxy, such as $^{10}$Be and $^{14}$C data of 11,000 years \citep{Uso23, Biswas23}.
Observational studies suggest a variation in the large-scale surface magnetic field of stars with its age which is known as `magnetochronology' \citep{Viddo14}. 
Moreover, studies \citep{Skumanich72, R84, Noyes84a} show that the younger stars rotate faster
and show much more magnetic activity than older stars.
Thus, the rotation rate (or period) largely controls the magnetic activity of Sun and Sun-like stars which is connected with their ages. 

Therefore, there must be a critical age or
rotation period for stars beyond which the generation of large-scale magnetic fields ceases. 
We note that the large-scale magnetic field is produced above the critical dynamo transition and, in some cases, in subcritical
regimes \citep{KO10, KKB15, Vindya21}.
However, the appealing question is, `at what regime is our solar dynamo operating, or in other words, how far is our Sun from the critical dynamo transition?
Observational study of \citet{Met16} reveals that the solar dynamo may be in the transition region of magnetically active and inactive branches of stars.
Theoretical modelings \citep{CS17, CS19, BT21} indicate that the solar dynamo is possibly near the transition to a magnetically inactive state.
Furthermore, \citet{KN17} 
showed that the solar dynamo is only about $10\%$ supercritical. Thus, this indicates that the solar dynamo is operating near the critical dynamo transition.  
Recently, \citet{Ghosh24} showed that the solar dynamo is not highly supercritical and operating near the dynamo's critical regime based on the nonlinear characterization of solar magnetic
cycles.
In the present manuscript, by conducting independent studies, we provide additional support that the solar dynamo operates only in the weakly supercritical regime.

The large-scale magnetic fields of the Sun and Sun-like stars are
believed to be the manifestation of the dynamo process operating in
their convection zones (CZs) \citep{Mof78, Pa55}.
In the $\alpha$ $\Omega$-type dynamo, which applies for the Sun, the important controlling parameter is the dynamo number $D$. The magnetic field grows when $D$ exceeds a critical value. Here, $D$ can be  defined as
\begin{equation}
%D = \frac{\alpha_0 \Delta \Omega R_\odot^3}{\eta_0^2}
D = \frac{\alpha_0 \Delta \Omega L^3}{\eta_0^2}
\label{eq0},
\end{equation}
where $\Delta \Omega $ is angular velocity
variation within CZ, 
$\alpha_0$ is the measure of the $\alpha$ effect, $L$ is the depth of CZ, and $\eta_0$ is the turbulent magnetic diffusivity.
In the classical $\alpha$ $\Omega$ type dynamo, helical convective $\alpha$ generates the poloidal field from the toroidal field \citep[e.g,][]{KR80}. However, in recent years, it has been realized that the generation of the poloidal field takes place due to the decay and dispersal of the tilted bipolar magnetic regions known as \bl\ mechanism \citep{Ba61, Leighton69},
which is strongly supported by observations \citep{Das10, KO11, Mord22, CS15, CS23}. 
\bl\ type models have emerged as a popular paradigm for explaining various features of regular solar magnetic cycle including the polarity reversals, poleward migration of surface magnetic field, and the equatorward migration of toroidal field \citep{Kar14a, Cha20, CS23}. 
As in this model, the decay of tilted bipolar magnetic regions (BMR) generates the poloidal field on the solar surface, we have a good observational back up of this part of the model. By monitoring the build-up rate of the polar field after its reversal \citep{Kumar22, BKK23} or by feeding the peak polar field information into a dynamo model \citep[e.g.,][]{Sch78, CCJ07, BN18, HC19, Kumar21}, one can predict the amplitude of the next solar cycle. 
One good thing about these models is that
the fluctuations in the BMR properties are observed and thus quantified \citep{JCS14, OCK13, Kumar24, Sreedevi24}. These fluctuations in the \bl\  dynamo models explain majority of the irregular features of solar cycle, including long-term modulation and grand minima \citep{CD00, Cha04, CK09, CK12, Kar10, OK13, Pas14, KM17, KM18}. 
Recently, \bl\ dynamo model has also been used to explore the features of stellar magnetic variability  \citep{V23}.

{{In recent years, global MHD simulations are producing encouraging results of the solar magnetic field and some large-scale flows such as solar-like differential rotation and meridional flow in some parameter regimes \citep[e.g.,][]{brun04, racine11, FF14, Kar15, ABMT15, Kap16, brun17, KMB18b, Strugarek18, brun22, Hotta22}. However, these simulations are still far from producing many features of surface magnetic field and deep convection \citep{brun17review, kapyla_ISSIRev}. Despite these, one still could employ global MHD simulation to identify the supercriticality of the Sun as the flows are self-consistently generated in the model and all nonlinearities are by default captured. However, we are not doing this because of two reasons. The first is because we need to run these simulations for several thousands of years to analyse the long-term modulation and grand minima, which is notoriously expensive numerically, and the second is due to difficulties in changing the dynamo number by keeping the critical dynamo transition point unaltered in global simulations. }}

Previous studies demonstrate that the variability of the magnetic cycle depends on the supercriticality of the dynamo 
\citep{CSZ05, KKB15, Cha20, Karak23}. 
Thus, changes in the supercriticality of the dynamo,
will cause change in the features of the solar cycle, including the properties of grand minima and maxima and the memory of the polar field \citep{kumar21b}.
Therefore, in this manuscript, we have
estimated the solar dynamo supercriticality
by analysing various features of magnetic cycle and comparing them with observations.
To do so, we have used several dynamo models, namely the \bl\ dynamo models used in \citet{kumar21b} and \citet{Pas14} and the time delay models of \citet{Ha14} and \citet{albert21}.
We identify the regime of solar dynamo operation by
estimating 
the recovery rate of Maunder Minimum (Section 3.1) and statistics of grand minima and maxima (Section 3.2).
%,and modelling last five solar cycles using the observed dipole moment (Section 3.3).
We conclude that the features of solar cycle favour weakly supercritical solar dynamo. 
We also make comments on whether the linear correlation between the available proxy of the polar field and next cycle amplitudes (Section 3.3.1) and the Gnevyshev-Ohl rule (Section 3.3.2) can be used to offer further support for our conclusion.

\section{Models}

We use various \bl\ type flux transport and time delay dynamo models to conduct this study.
Below we briefly discussed these models.

\subsection{Flux transport dynamos}

For the flux transport dynamo model \citep{CSD95, Kar14a, Hazra23}, we solve the following equations for axisymmetric magnetic field:

\begin{equation}\label{eq1}
\frac{\partial A}{\partial t} + \frac{1}{s}(\vec{v}\cdot{\bf \nabla})(s A)
 = \eta_{p} \left( \nabla^2 - \frac{1}{s^2} \right) A + S_{\alpha},
\end{equation}

\begin{equation}
\label{eq2}
%    \begin{split}
\frac{\partial B}{\partial t}+\frac{1}{r}\left[\frac{\partial \left(rv_rB\right)}{\partial r} + \frac{\partial \left(v_{\theta}B\right)}{\partial\theta}\right]=\eta_t\left(\nabla^2-\frac{1}{s^2}\right)B + s\left( \vec{B}_p.\nabla\right)\Omega \\
 + \frac{1}{r}\frac{d\eta_t}{dr}\frac{\partial}{\partial r}\left(rB\right),
%    \end{split}
\end{equation}
where $A$ is the potential of poloidal magnetic field and $B$ is the toroidal magnetic field, 
$s = r\sin{\theta}$ with $\theta$ being colatitude, $\vec{v}=v_r {\bf \hat{e}_r} + v_{\theta} {\bf \hat{e}_{\theta}}$
is the meridional circulation, $\Omega$ is the angular velocity, $\eta_p$ and $\eta_t$ are the turbulent diffusivities of the
poloidal and toroidal fields, respectively, $\alpha$ is the parameter that captures the \bl\ process for the generation of the poloidal field from the toroidal one. 
In this study, we use three \bl\ models, namely Models~I, II, and III.

%In this study, we used the same parameters and models 
% I, II, 
%III, and IV \red{\textbf{(here we call them Model I and II)}}, which were used in \citet{kumar21b}.
% For Model~I and II we use local $\alpha$ prescription of the form $S_{\alpha}= \alpha B$ where, 
% \begin{equation}
% \alpha = \frac{\alpha_0}{4}\cos\theta \left[1 + \mathrm{erf} \left(\frac{r - 0.95R_\odot}{0.025R_\odot}\right) \right]
% \times \left[1 - \mathrm{erf} \left(\frac{r - R_\odot}{0.025R_\odot}\right) \right].
% \label{eq:alpha}
% \end{equation}

% Only difference between Model~I and II is the amplitude of the meridional flow and it is 15~\mps\ for Model I and 26~\mps\ for Model II.
% \blue{\textbf{
% We note that in Models I and II, two different values for the diffusivity for poloidal and toroidal fields are taken. The justification for this is that the strong toroidal field remains confined in localized regions and experiences more quenching compared to the poloidal one, which remains spread \citep[see][where this idea was introduced]{CNC04}.  
% }}

For Models~I and II we use a nonlocal $\alpha$, which was used in earlier studies \citep{DC99, CNC05, CH16}.
Thus, in these models,
\begin{equation}
S_{\alpha}= \frac{\alpha}{1+\left(\frac{B(0.7R_\odot,\theta)}{B_{0}}\right)^2}B(0.7R_\odot,\theta),
\end{equation}
where $\alpha$ is given by 
\begin{equation}
%\begin{split}
\alpha=\frac{\alpha_0}{4}
\sin\theta\cos\theta\left[\frac{1}{1+e^{-\gamma(\theta-\frac{\pi}{4})}}\right] 
 \left[1+\mathrm{erf}\left(\frac{r-0.95R_\odot}{0.05R_\odot}\right)\right]  \left[1-\mathrm{erf}\left(\frac{r-R_\odot}{0.01R_\odot}\right)\right]
\label{alphaprof_nonlocal}
%\end{split}
\end{equation}
where $\gamma=30$
{{and $\alpha_0$ is the amplitude of $\alpha$ effect.
When we introduce fluctuations in these models, we multiply $\alpha_0$ by a Gaussian of unity mean and 2.67 standard deviation as inspired by the study of  \citet{OCK13} using sunspot Catalog of Solar Activity
of the Pulkovo Observatory.
%replace $\alpha_0$ by $\alpha_0F$, where $F$ is the Gaussian fluctuation with mean =1, and $\sigma_d = 2.67$ respectively.
}}
Also, for diffusivity, in these models, we consider $\eta_t=\eta_p = \eta$, where 
\newcommand{\etaRZ}{\eta_{\mathrm{RZ}}}
\newcommand{\etasurf}{\eta_{\mathrm{surf}}}
\def\Rs{R_{\odot}}
\begin{equation}
\eta(r) = \eta_{RZ} + \frac{\eta_0}{2}\left[1 + \mathrm{erf} \left(\frac{r - 0.7\Rs}
{0.02\Rs}\right) \right]
+\frac{\etasurf}{2}\left[1 + \mathrm{erf} \left(\frac{r - 0.9\Rs}
{0.02\Rs}\right) \right],
\label{eq:eta}
\end{equation}
where $\eta_{RZ} = 5\times10^{8}$~\cmss\ and
$\etasurf =  2\times10^{12}$~\cmss.
For {{Model~I, we have considered $\eta_0 =  5\times10^{10}$~\cmss,
and for Model~II, we have used five times less diffusivity than in Model~I.}} 

Besides these models, we have also used the model of \citet{Pas14} in which a weak mean-field $\alpha$ effect \citep{Pa55} is included in addition to the usual \bl\ source for the poloidal field (local prescription for $\alpha$) and total poloidal field source considered as $S_\alpha = \alpha B$. Hence the total  $\alpha$ effect in this model is defined as $\alpha = \alpha_{BL} + \alpha_{MF}$. Here, 
\begin{equation}
\begin{split}
\alpha_{BL} = \alpha_{0BL}\frac{\cos\theta}{4} \left[1 + \mathrm{erf} \left(\frac{r - 0.95R_\odot}{0.025R_\odot}\right) \right]
 \left[1 - \mathrm{erf} \left(\frac{r - R_\odot}{0.025R_\odot}\right) \right] \\ \times a_1 \left[1 + {\rm erf}\left(\frac{B_{\phi}^2 - B_{1lo}^2}{d_3^2}\right)\right] \left[1 - {\rm erf}\left(\frac{B_{\phi}^2 - B_{1up}^2}{d_4^2}\right)\right],
\end{split}
\end{equation}
with $B_{1lo} = 10^3$~G, $B_{1up} = 10^5$~G, $d_3 = 10^2$~G, and $d_4 = 10^6$~G and $a_1 = 0.393$ is the normalization constant. 
{{$\alpha_{0BL}$ determines the amplitude of the \bl\ $\alpha$ coefficient, whose value is varied to change the supercriticality of the dynamo.
Note that in addition to the upper quenching, there is a lower quenching of \bl\ effect, which becomes important when the toroidal field falls below $10^3$~G in this model.
This lower cut-off may not be well justified because BMRs do not show any cut-off in the field strength, and even the small BMRs show some systematic tilt, although the scatter around Joy's law increases with the decrease of flux in the BMR \citep{jha20, Sreedevi24}.
However, as we are using the model of \cite{Pas14} as it is, we are following the same procedure of lower quenching.
Then, this model also includes a mean-field $\alpha$, which operates at the weak field regime. 
As argued by \cite{Pas14}, when the magnetic field becomes weak during grand minima and \bl\ $\alpha$ is inefficient, this additional $\alpha$ helps recover the model from grand minima. The mean-field $\alpha$ is given by,}}
\begin{equation}
%\begin{split}
\alpha_{\rm MF} = \alpha_{\rm 0MF} \frac{\cos\theta}{4} \left[1 + \mathrm{erf} \left(\frac{r - r_1}{r_2}\right) \right] 
\left[1 - \mathrm{erf} \left(\frac{r - R_\odot}{r_2}\right) \right]  \frac{1}{\left[1 + \left(\frac{B_\phi}{B_0}\right)^2\right]},
%\end{split}
\end{equation}
with 
{{$\alpha_{\rm 0MF} = 0.4$, which determines the amplitude of the mean-field $\alpha$, $r_1 = 0.713R_\odot$, $r_2 = 0.025R_\odot$, and $B_0 = 10^4$~G.}}
{{We introduce stochastic fluctuations in this model by replacing $\alpha$ with $\alpha(1 + f\sigma(t, \tau_{\text{cor}}))$, where $\sigma$ is a uniform random deviation within [$-1$, $1$] interval, $\tau_{\text{cor}}$ is the time interval after which $\alpha_0$ is updated, and $f$ is the percentage level of fluctuations.}}
% This is the similar way as in model I and model II.}}

{{We note that in this model, different values of diffusivity for the poloidal ($\eta_p$) and the toroidal field ($\eta_t$) are used; see \Fig{fig:profile}, right panel. The toroidal field diffusivity in the bulk of the CZ is about 50 times smaller than that of poloidal field. The justification for this is that the strong toroidal field remains confined in localized regions and experiences more quenching compared to the poloidal one, which remains spread \citep[see][where this idea was introduced]{CNC04}.}}

The \mc\ profile of this model is the same as used in the previous models, except the parameter $v_0$ whose value is taken to be $-29$~\mps\ in this model.
For the other details of this model, we refer readers to \citet{Pas14}. 
We call this model as {{Model III}}.
{{
We note that for all the models, the angular frequency $\Omega$ is taken from 
an analytic approximation of the helioseismic data used in most kinematic dynamo models, precisely Equation 4 of \citet{DC99}.
}}
Finally, we have used two other models, namely {{Models IV and V}}, which are time delay dynamo models as described below. 
%\red{We note that the previous studies simulations set-up with fluctuations in a dynamo system with \bl\ mechanism as a poloidal field source can alone recover from a grand minimum phase \citep{CD00, CK12, KM18}. However, \citet{Ch05, Ha14} studies raise questions about solar cycle recovery from grand minima through \bl\ mechanism due to the physics behind its mechanism. They suggested that one should ensure that the generation of the polar field is only due to buoyantly erupted active regions through the \bl\ mechanism and not due to a weaker field, which may reach the surface due to advection or diffusion. Moreover, they suggested a lower cut-off in $alpha_{BL}$ to get a buoyantly erupted active region on the solar surface due to a strong magnetic field. Furthermore, using \bl\ dynamo, \citet{Ha14} could not recover from the grand minimum phase. Therefore, to bypass this problem, \citet{Pas14} added a weak mean field alpha effect with the \bl\ mechanism. Thus, in this study, we employ both the \bl\ mechanism alone and the \bl\ mechanism with a mean-field alpha effect to study the grand minima and its recovery.}
%%%%%%%%%%%%%%%%%%%%%%%%%%%%%%%%%%%%%%%%%%%%%%%%%%

\subsection{Time delay dynamo}
In time delay dynamo model \citep{wilsmith},
we solve the following two truncated (removing all spatial dependence and taking into account the spatial segregation in the source regions) equations,
\begin{equation}\label{eq6}
    \frac{dB(t)}{dt} = \frac{\omega}{L} A(t - T_0) - \frac{B(t)}{\tau_d},
\end{equation}
\begin{equation}\label{eq7}
    \frac{dA(t)}{dt} = \alpha_0 f_0(B(t-T_1))B(t-T_1) - \frac{A(t)}{\tau_d},
\end{equation}

where $\omega$ and $L$ represent the differential rotation and length scale in the tachocline, respectively, $\tau_d$ denotes the diffusion timescale of the turbulent diffusion in the CZ, 
while $\alpha_0$ is the amplitude of the $\alpha$-effect, similar to those used in flux-transport dynamo models. 
The parameters $T_0$ and $T_1$ account for the time delay in the conversion of the poloidal field into the toroidal field and vice versa. The factor $f_0$ is a quenching factor, which is approximated here by a nonlinear function:

\begin{equation}\label{eq8}
    f_0 = \frac{1}{4}\left[1 + {\rm erf}\left({B^2(t) - B^2_{min}}\right)\right]\left[1 - {\rm erf}\left({B^2(t) - B^2_{max}}\right)\right],
\end{equation}

where $B_{max}$ and $B_{min}$ are the upper and lower limit to the toroidal field strength between which the $\alpha$-effect can act (same as the previous model).
% Parameters $B_{min} = 1$, and $B_{max} = 10$ are used for quenching.

{{
For Model IV, we follow \citet{Ha14} with additional weak mean field poloidal source in the equation \eq{eq7} similar to Model III. Thus equation \eq{eq7} modified as,
}}

\begin{equation}\label{eq13}
    \frac{dA(t)}{dt} = \alpha_0 f_{0}(B(t-T_1))B(t-T_1)  + \alpha_{mf}f_1(B(t-T_2))B(t-T_2)  - \frac{A(t)}{\tau_d},
\end{equation}

{{where $\alpha_{mf} = 0.2$, is the mean-field poloidal source and $f_1$ is the corresponding quenching which is given by,
\begin{equation}\label{14}
    f_1 = \frac{1 - {\rm erf}\left(B^2(t-T_2) - B_{eq}^2\right)}{2}
\end{equation}
with $B^2_{eq} = 1$ and $T_2 = 0.25$, is the time delay that is necessary for the toroidal field to enter the source region where the additional weak-field $\alpha$ effect is located.
The variation of the quenching profile $f_0$ and $f_1$ is shown in \Fig{fig:quench}.
We take $\tau_d = 15$, $B_{min} = 1$, $B_{max} = 7$, $T_1 = 0.5$, $T_0 = 2$, and $\omega/L = -0.34$ for this model}}. 

{{Finally for Model V, we follow \citet{albert21}, 
and take following values of the parameters: 
$\alpha_{mf} = 0$ (no mean-field $\alpha$), $B_{min} = 1$, $B_{max}=10$, $\tau_d = 1$, and $(T_0 + T1) / \tau_d = 0.82$. In this model, the quenching profile is similar to $f_0$ except for the upper and lower cut-off values of the magnetic field and is presented as $f'$ in \Fig{fig:quench} in the Appendix.}}
{{For these two models (IV and V), we incorporate stochastic fluctuations in $\alpha_0$ and $\alpha_{mf}$ in the same way as in Model III.}}

%%%%%%%%%%%%%%%%%%%%%%%%%%%%%%%%%%%%%%%%%%%%%%%%%%
% \section{Methods}
%%%%%%%%%%%%%%%%%%%%%%%%%%%%%%%%%%%%%%%%%%%%%%%%%%

\section{Results and Discussion}

In all the dynamo models used in our studies, the amplitude of the $\alpha$ coefficient, $\alpha_0$, determines the dynamo efficiency. Growing magnetic field, i.e., dynamo action, is possible only when the value of $\alpha_0$ exceeds a certain value, which is defined as $\alpha_0^{crit}$. Its value is obviously different for different models, which is given in the first column of Table~1.  
Thus, we define a quantity, $\hat{\alpha}_0 = \alpha_0 /  \alpha_0^{crit}$, which can be a measure of the amount of dynamo supercriticality. As we increase the value of $\hat{\alpha}_0$ beyond unity, the model becomes more and more supercritical. We note that this $\hat{\alpha}_0$ is not exactly the dynamo number,  which can have complicated dependence on parameters in the present flux transport dynamo models (unlike Eq. 1 for simple $\alpha$ $\Omega$ dynamo model) but it is related to it. In our study, for each model, we perform a set of solar cycle simulations at  $\hat{\alpha}_0 =2$, 4, and 8, and for each model, the level of fluctuations in $\alpha$ remains the same.

From each model, we compute several properties, namely, the recovery rate from Maunder-like grand minimum, the number and duration of grand minimum and maximum, 
%the accuracy of cycle prediction using the polar field,
the correlation between the polar field at cycle minimum with the amplitude of the next cycle toroidal field, and the Gnevyshev–Ohl/Even–Odd Rule. Below, we discuss their results.

\begin{figure*}
\centering
\includegraphics[width=\linewidth]{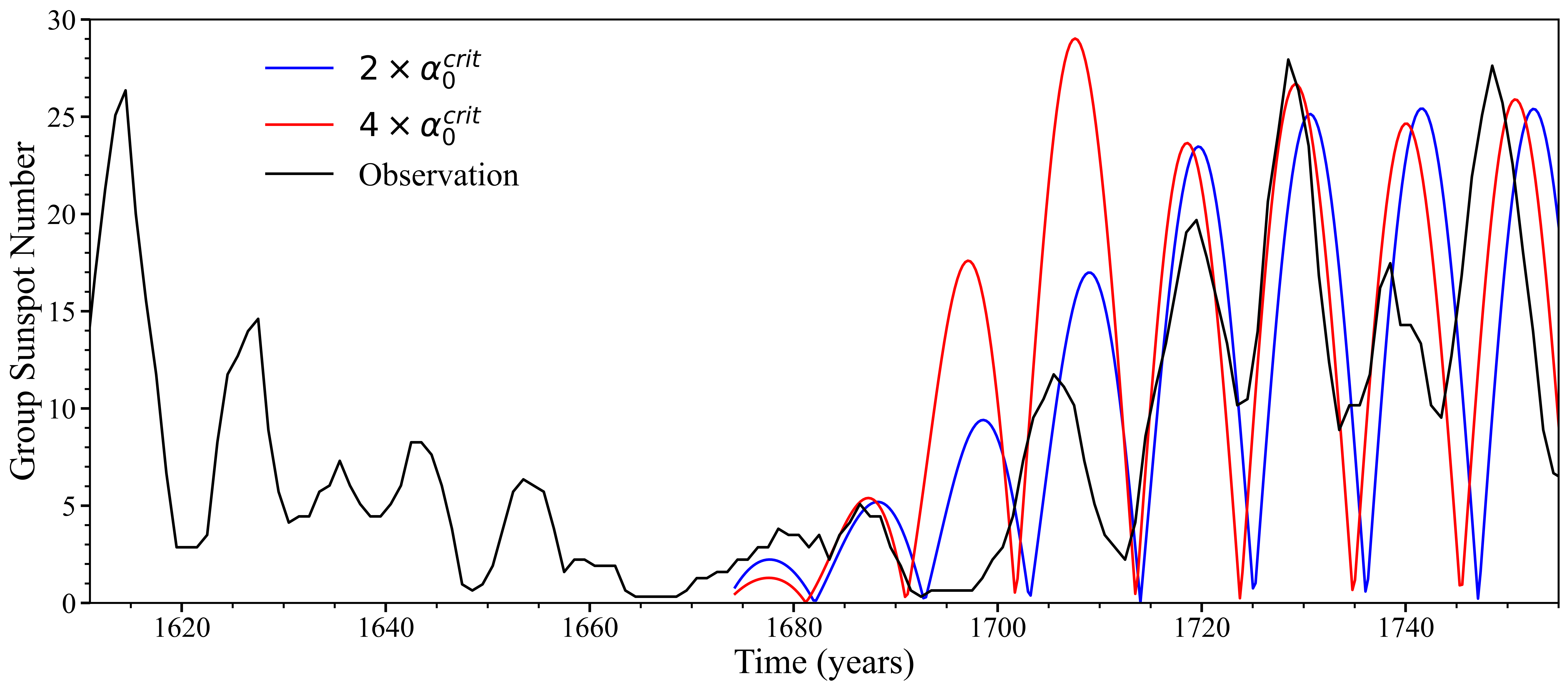}
\caption{The plot depicting the comparison of recovery rates from Maunder minimum. The black line is the observational data of Maunder minimum (yearly mean group sunspot number available during 1610–2015 obtained from WDC-SILSO, Royal Observatory of Belgium, Brussels). The blue and red lines represent the model data at  $\hat{\alpha}_0=2$, and $\hat{\alpha}_0=4$ from {{Model I}}.}
\label{fig:fig1}
\end{figure*} 

\begin{figure}
\centering
\includegraphics[width=\linewidth]{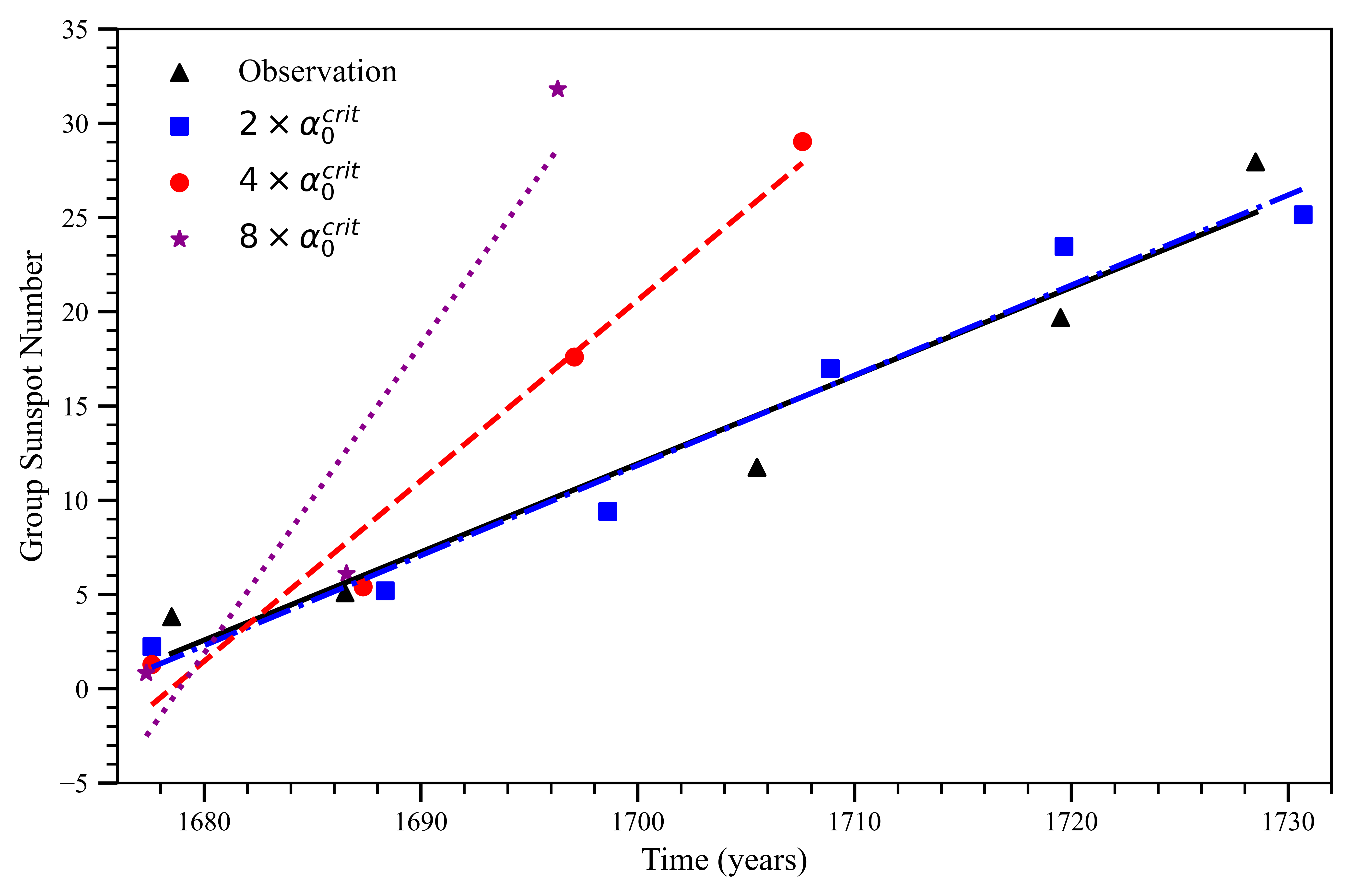}
\caption{The plot compares the recovery rates from the Maunder Minimum. Black/triangles show the observed recovery rate, and blue/squares, red/circles, and dark magenta/asterisks (and connecting lines)  represent model recovery rates at $\hat{\alpha}_0=2$, 4, and 8 respectively in {{Model I}}.}
\label{fig:fig2p1}
\end{figure} 

\subsection{Recovery Rate of Maunder Minimum}
We know that during Maunder minimum, the Sun went to a deep minimum of magnetic activity for at least 40 years \citep{Eddy, RN93, Uso23}. 
While the triggering phase of the Maunder minimum is somewhat uncertain, its recovery phase is a bit gradual. During about 1700 to 1730, the Sun gradually recovered from the quiet phase to the normal one.  Although we have these details for only one grand minimum, we can use this rate of recovery from the Maunder minimum to learn about the dynamo growth rate, which is related to the supercriticality of the Sun.  To do so, we manually cut the magnetic field in the dynamo model by multiplying the poloidal field at the solar minimum by a factor of $\gamma_p = 0.1$.  We note that this idea was based on the work of \citet{CK09}, who proposed that due to fluctuations in the Babcock-Leighton process, the polar field at the solar minimum can be largely different, and to capture this in their model, they multiplied the polar field at the solar minimum by a factor  $\gamma_p$.  After multiplying the polar field at the solar minimum by a factor of $\gamma_p$, we run the model for several cycles to examine its recovery to the normal phase at different values of $\hat{\alpha}_0$.

\Fig{fig:fig1} shows the recovery of the model from the  Maunder minimum in {{Models I}} at $\hat{\alpha}_0 = 2$ and 4 (blue and red curves) and their comparison with the observed data (black curve).  

To make a better comparison of the observed peaks of the cycles during the recovery phase of the Maunder minimum with the model, in \Fig{fig:fig2p1} we present the slopes of the observed data from the Maunder minimum (black curve) and from the model (blue, red and dark magenta) at $\hat{\alpha}_0 = 2$, 4 and 8. 
We see that as $\hat{\alpha}_0$ is increased, the model deviates more and more from the observed one. {{We have repeated this exercise using other models and find consistent results in Models II, III and IV. The Model V is however excluded in this study because this model produces chaotic solution and the recovery rate is quite different even at a slight different value of $\alpha_0$. }} 

The reason for the robust result in all four models (I--VI) is not difficult to understand. As $\hat{\alpha}_0$ increases, the dynamo becomes stronger (dynamo number increases), and it allows the magnetic field to grow more rapidly. 
In summary, from the analysis of the recovery rate of Maunder minimum, we 
can expect that the supercriticality of the sun as measured by  $\hat{\alpha}_0$ is around two.

\begin{table}
% \centering
\caption{
{{ 
The number of grand minima and maxima obtained from 11,000 years of simulations using different models at different dynamo supercriticality as measured by $\hat{\alpha}_0$. 
}}
The critical value for the \bl\ $\alpha$ in each model is given in the second column.}
\begin{tabular}{c|c|c c c|c c c}
\hline
Models & $\alpha^{\rm crit}_0$& \multicolumn{3}{c}{Grand Minima} & \multicolumn{3}{c}{Grand Maxima} \\ 
($\hat{\alpha}_0$) & & 2.0 & 4.0 & 8.0 & 2.0 & 4.0 & 8.0\\ \hline
%{Model I} \\ \citet{kumar21b} & 0 & 0 & 0 & 0 & 0 & 0 \\ \hline
%{Model II} \\ \citet{kumar21b} & - & 3 & 0 & - & 0 & 0 \\ \hline
{{{Model I}}} & & & & & & & \\ \citet{kumar21b} & 0.38 & 20 & 8 & 5 & 22 & 16 & 18 \\($\sigma=2.67$ Gaussian fluctuation) & (m s$^{-1}$) & & & & & & \\ \hline
{{{Model II}}} & & & & & & & \\ \citet{kumar21b}& 0.05 & 18 & 8 & 6 & 22 & 15 & 13 \\($\sigma=2.67$ Gaussian fluctuations) & (m s$^{-1}$) & & & & & & \\ \hline
{{{Model III}}} & & & & & & & \\ \citet{Pas14} & 18.9 & 16 & 9 & 4 & 26 & 8 & 3 \\(200\% uniform fluctuations) & (m s$^{-1}$) & & & & & & \\ \hline
{{{Model IV}}} & & & & & & & \\ \citet{Ha14} & 0.16 & 26 & 11 & 10 & 27 & 21 & 21 \\(100\% uniform fluctuations) & & & & & & & \\ \hline
{{{Model V}}} & & & & & & & \\ \citet{albert21} & 10.2 & 20 & 16 & 9 & 24 & 14 & 9 \\(70\% uniform fluctuations) & & & & & & & \\ \hline
\multicolumn{2}{c}{Observed~~~~~~~~~~~} & \multicolumn{3}{c}{27} & \multicolumn{3}{c}{23} \\ \hline
\end{tabular}
\label{tab1}
\end{table}

\begin{table}
% \centering
\caption{
Percentage of time spent in grand minima and maxima obtained for different values of $\hat{\alpha}_0$. The values in parentheses are the average duration of grand minima and maxima in years. For the time delay models, the time is in a dimensionless unit and thus not shown. 
}
\begin{tabular}{c|c c c|c c c}
\hline
Models & \multicolumn{3}{c}{Minima} & \multicolumn{3}{c}{Maxima} \\ 
($\hat{\alpha}_0$) & 2.0 & 4.0 & 8.0 & 2.0 & 4.0 & 8.0\\ \hline
%{Model I} \\ \citet{kumar21b} & 0 & 0 & 0 & 0 & 0 & 0 \\ \hline 
%{Model II} \\ \citet{kumar21b} & - & 3 & 0 & - & 0 & 0 \\ \hline
{{{Model I}}} & 6.0 & 2.3 & 1.2 & 6.8 & 4.8 & 5.7 \\
\citet{kumar21b} & (33.9) & (31) & (24.4) & (34.8) & (33.5) & (33.4) \\($\sigma=2.67$ Gaussian fluctuation) & & & & & & \\ \hline
{{{Model II}}} & 4.3 & 1.8 & 1.7 & 7.2 & 4.8 & 4.0 \\
\citet{kumar21b} & (29.6) & (26.6) & (31.3) & (36.6) & (33.9) & (33) \\($\sigma=2.67$ Gaussian fluctuation) & & & & & & \\ \hline
{{{Model III}}}  & 13.4 & 3.4 & 0.8 & 13.2 & 2.3 & 0.74 \\\citet{Pas14} & (92.1) & (41.1) & (22.5) & (55.8) & (31.3) & (27) \\ (200\% fluctuation) & & & & & & \\ \hline
{{{Model IV}}} & 5.4 & 0.4 & 0.4 & 4.9 & 0.9 & 0.6\\ \citet{Ha14} &  &  &  &  &  &  \\ (100\% fluctuation) & & & & & & \\ \hline
{{{Model V}}} & 22.5 & 30.0 & 7.5 & 41.7 & 22.3 & 10.5 \\\citet{albert21}  &  &  &  &  &  &  \\ (70\% fluctuation) & & & & & & \\ \hline
\end{tabular}
\label{tab2}
\end{table}

\begin{figure*}

\includegraphics[width=\linewidth]{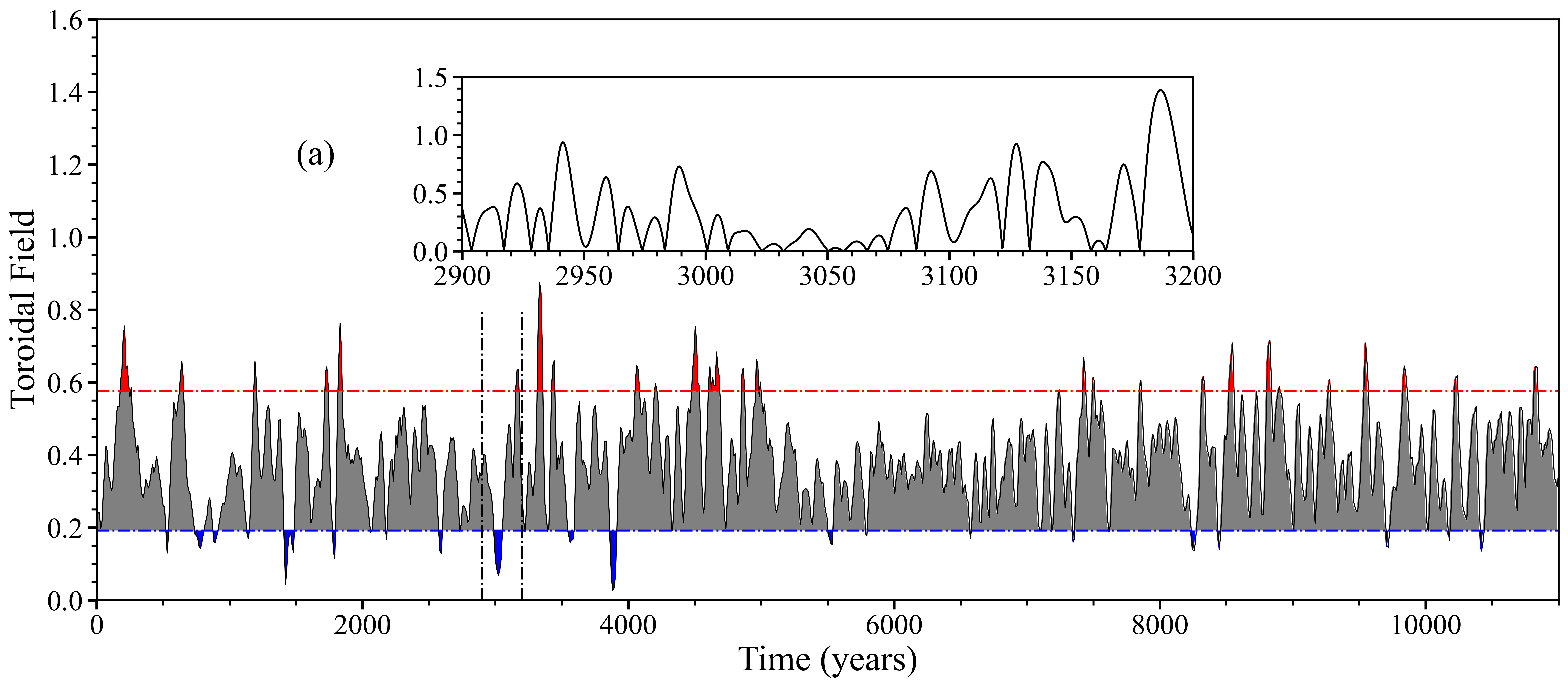}
\includegraphics[width=\linewidth]{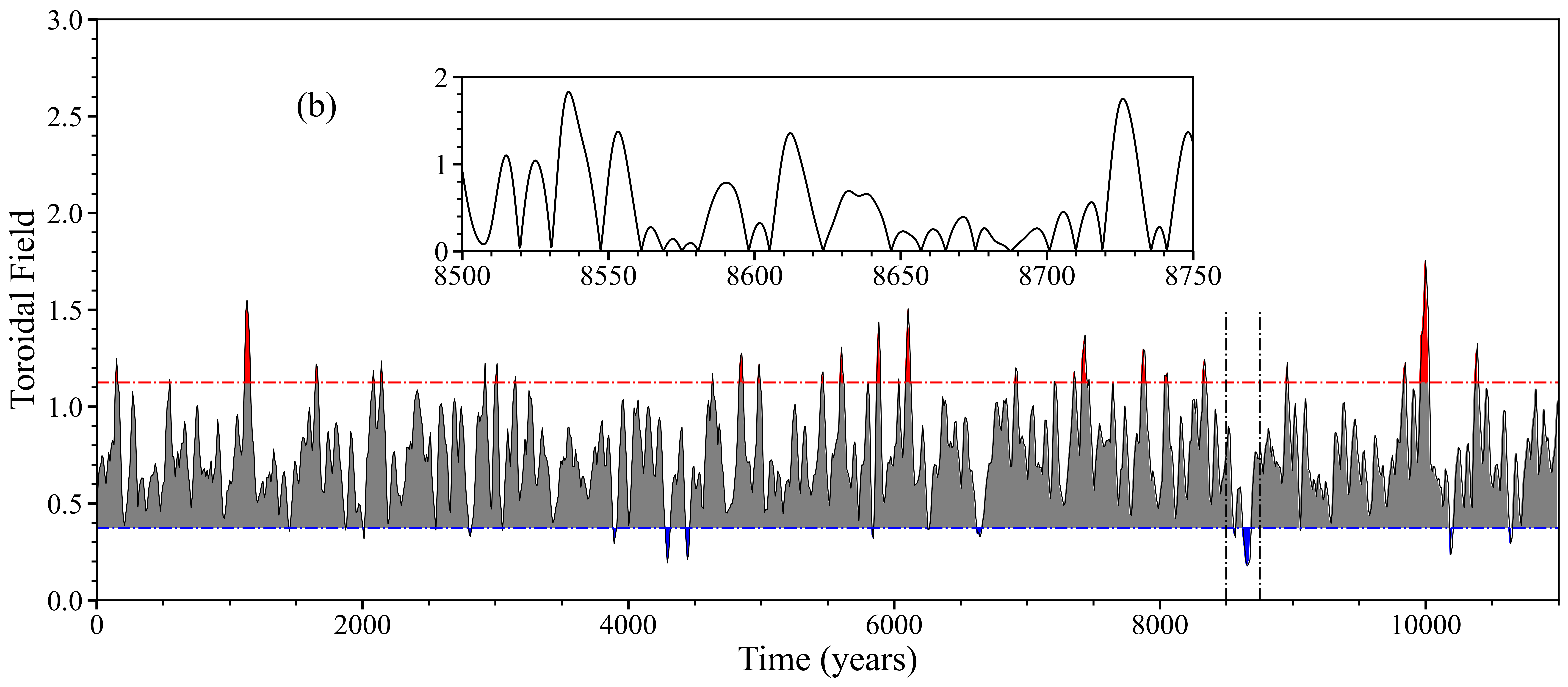}
\caption{Temporal variation of the smoothed toroidal field of 11,000-year simulation of {{Model I}}. Blue-shaded regions below the horizontal blue line represent the grand minima, whereas red-shaded regions above the horizontal red line represent the grand maxima; (a) for $\hat{\alpha}_0=2$, and (b) for $\hat{\alpha}_0=4$.
Insets show epochs around a grand minimum}
\label{fig:fig2}

\end{figure*}

%%%%%%%%%%%%%%%%%%%%%%%%%%%%%%%%%%%%%%%%%%%%%%%%%%

\subsection{Statistics of Grand Minima and Maxima}

Next, we compute the grand minima and maxima frequency using different models with varying supercriticality.
To produce grand minima and maxima in our kinematic dynamo, we follow the traditional methods, i.e., we include fluctuations in the poloidal source to mimic the variations in the flux emergence and BMR properties \citep{Karak23}.  Explicitly, we include stochastic fluctuations in the ampolitude of the $\alpha$ (both in \bl\ and mean-field) as mentioned in the Section 2.
%/$(1 + (B_0/B)^2)$, where $B_0$ is the initial magnetic field.
The details of the level of fluctuations and the coherence time in each model are mentioned in the first column of Table 1 (also see model section). 
By running five models at different values of $\hat{\alpha}_0$, each for 11,000 years (the same duration for which the reconstructed solar activity is available; \citet{Uso23}), 
we compute the statistics of grand minima and maxima.  To compute the grand minima/maxima, we follow the same procedure as used in the observational study of \citet{USK07}.  We first bin the toroidal flux data computed at the base of CZ at low latitude ($r=0.72R_\odot$, $\theta = 10^{\circ}$ to $45^{\circ}$) 
using a window of the length of the average cycle duration and then smoothing the binned data using Gleissberg’s low-pass filter 
 {{ 
1-2-2-2-1. 
}}
The time series of these data is shown in \Fig{fig:fig2} for $\hat{\alpha}_0 = 2$ and 4 from  {{Model I}}. 
 Finally, we define an episode as a grand minimum (maximum) when this smoothed data remains below (or above) $50\%$ ($150\%$) of the average of the whole time series for at least two-cycle duration continuously; see blue and red shaded regions for grand minima and maxima in \Fig{fig:fig2}. 
 % \red{\textbf{It is important to note that}}
 % \blue{\textbf{  
 % we exclude Models I and II for this grand minima/maxima analysis because these models (with local $\alpha$ prescription) are prone to decay after they enter into a deep grand minimum. As already reported in the past \citep{KC13, Ha14}, a weak mean-field  $\alpha$ is beneficial to recover the model from the deep grand minimum.
 % While Models III, IV, and VII have no lower cut-off in the \bl\ $\alpha$ and Models V and VI include an additional mean-field $\alpha$. Thus, these models do not decay when they enter into a deep grand minima, and we consider these them for grand minima/maxima analyses.
 % }} 
As shown in \Tab{tab1}, at $\hat{\alpha}_0 = 2$, the numbers of grand minima in {{Model I and II}} are 20 and 18, respectively, and as 
the value of  $\hat{\alpha}_0$ is increased, the number decreases.
A similar result is seen for grand maxima, although the decrease is not monotonous.
To explore the robustness of these trends of grand minima/maxima with the $\hat{\alpha}_0$,
we repeat these studies using  other models, namely, Models III-VI. 
%\citep{Pas14} model and time delay models of  \citep{Ha14} and \citet{albert21}.  
We find similar trends in these models as well; see \Tab{tab1}. 
We recall that from the cosmogenic isotope records of \citet{USK07}, we know that 
Sun produced 27 grand minima and 23 grand maxima in the last 11,000 years.
Thus, it suggests that the models at  $\hat{\alpha}_0 = 2$ are closer to observations, 
implying that the solar dynamo is operating near the critical regime of the dynamo and far from the supercritical regime. 
Our result also agrees with the previous suggestions that the grand minima are more frequent in weakly supercritical dynamo \citep{KC13, OK13}.

\begin{figure*}
\centering
\includegraphics[width=\linewidth]{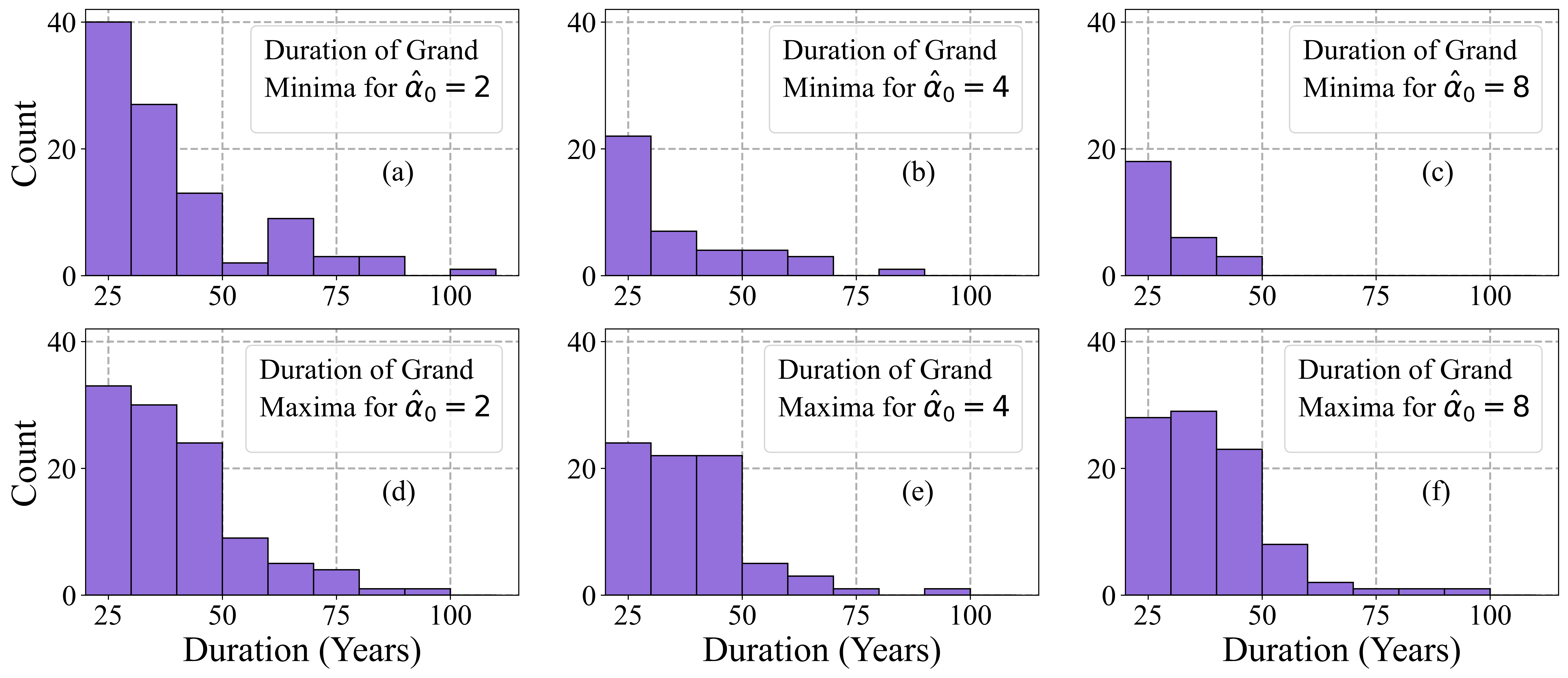}
\caption{{{Histograms for the durations}} of grand minima (a,b,c) and grand maxima (d, e, f) with increasing dynamo supercriticality ($\hat{\alpha_0}$).}
\label{fig:fig3}
\end{figure*}

Further analysing the properties of grand minima/maxima, we find that the percentage of time spent in grand minima decreases and the duration of grand minima becomes shorter with the increase of the supercriticality of the dynamo (\Fig{fig:fig3} and \Tab{tab2}). The long duration grand minima (Sp\"orer-like) are not seen at $\hat{\alpha}_0 = 4$. Even the Maunder-like grand minima are not detected at $\hat{\alpha}_0 = 8$; \Fig{fig:fig3}c.  For the grand maxima, surprisingly, we do not observe much change in the duration.  In all models considered in our study, we observe similar behaviour as demonstrated by the change of the average duration of grand minima/maxima with $\hat{\alpha}_0$. 
Again, from the average duration of grand minima and maxima at different $\hat{\alpha}_0$ (\Tab{tab2} and \Fig{fig:fig3}), we conclude that the $\hat{\alpha}_0 = 2$ is more favorable for the solar dynamo.

\subsection{Do the features of solar cycle at two times critical dynamo contradict with observations?}
\subsubsection{Correlation between the polar field and the sunspot cycle}

\begin{table*}
    % \centering
    \begin{tabular}{c|ccc}
    \hline
        $\phi_{r}(n)$ \& & Model I & Model II & Observations\\
        \hline
        $\phi_{tor}(n)$ & $-0.07$ & 0.19 & 0.16\\
        $\phi_{tor}(n+1)$ & 0.99 & 0.97 & 0.60\\
        $\phi_{tor}(n+2)$ & $-0.09$ & 0.29 & 0.27\\
        $\phi_{tor}(n+3)$ & 0.18 & 0.20 & -0.23\\
    \end{tabular}
    \caption{
    {{Comparison of the Pearson correlation coefficient between the polar flux (or its proxy) of cycle $n$ and toroidal flux (or sunspot area) of cycles $n$, $n+1$, $n+2$, and $n+3$ from Models I and II, and observation. The values from models are obtained from \citet{kumar21b} at 2 times critical ($\hat{\alpha_0} = 2$), while the observed values are taken from \citet{Muno13}}}.
    }
    \label{tab:corr}
\end{table*}

In any $\alpha$ $\Omega$ dynamo model, as long as the poloidal field acts as the source for the toroidal field, there exists a strong correlation between the poloidal (or polar) field at the cycle minimum and the toroidal field of the next cycle \citep{CB11}. Now the question is whether this memory of the poloidal field is propagated beyond one cycle toroidal field.  \citet{kumar21b} showed that the answer depends on the supercriticality of the dynamo.  When the dynamo operates near critical dynamo transition, both the magnetic field growth rate and the nonlinearity are weak, which supports multi cycle memory of the polar field.  On the other hand, when the dynamo is highly supercritical, the growth rate of the magnetic field is high, and the nonlinearity is also efficient, which tries to break the memory of the magnetic field.  By performing stochastically forced dynamo simulations at different parameter regimes and varying prescriptions for the poloidal sources,  \citet{kumar21b} demonstrated that the memory of the poloidal field is propagated to multiple cycles when the dynamo operates near the critical dynamo transition.  However, if the dynamo is highly supercritical, then the memory of the polar field is limited to only the next cycle toroidal field \citep[also see][for supporting this result using an independent study]{Ghosh24}.

Our present study favours $\hat{\alpha}_0=2$ for the Sun, which implies that the solar dynamo is neither critical nor heavily supercritical.  
{{Thus, following \citet{kumar21b} study, we expect a negligible correlation between the polar field at the end of cycle $n$ with the toroidal field for cycle $n+2$ and $n+3$; see \Tab{tab:corr} from \citet{kumar21b}). On the other hand, analysing the polar faculae counts--a proxy of the polar field--\citet{Muno13} showed that there is no significant correlation between the polar field (faculae count) of cycle $n$ with the toroidal field (sunspot area) of cycles $n+2$ and $n+3$; see the last column of \Tab{tab:corr}.  Hence, the conclusion that the Sun is about two times supercritical does not contradict the observed data.
}}

\begin{figure*}
    \centering
    \includegraphics[width=\linewidth]{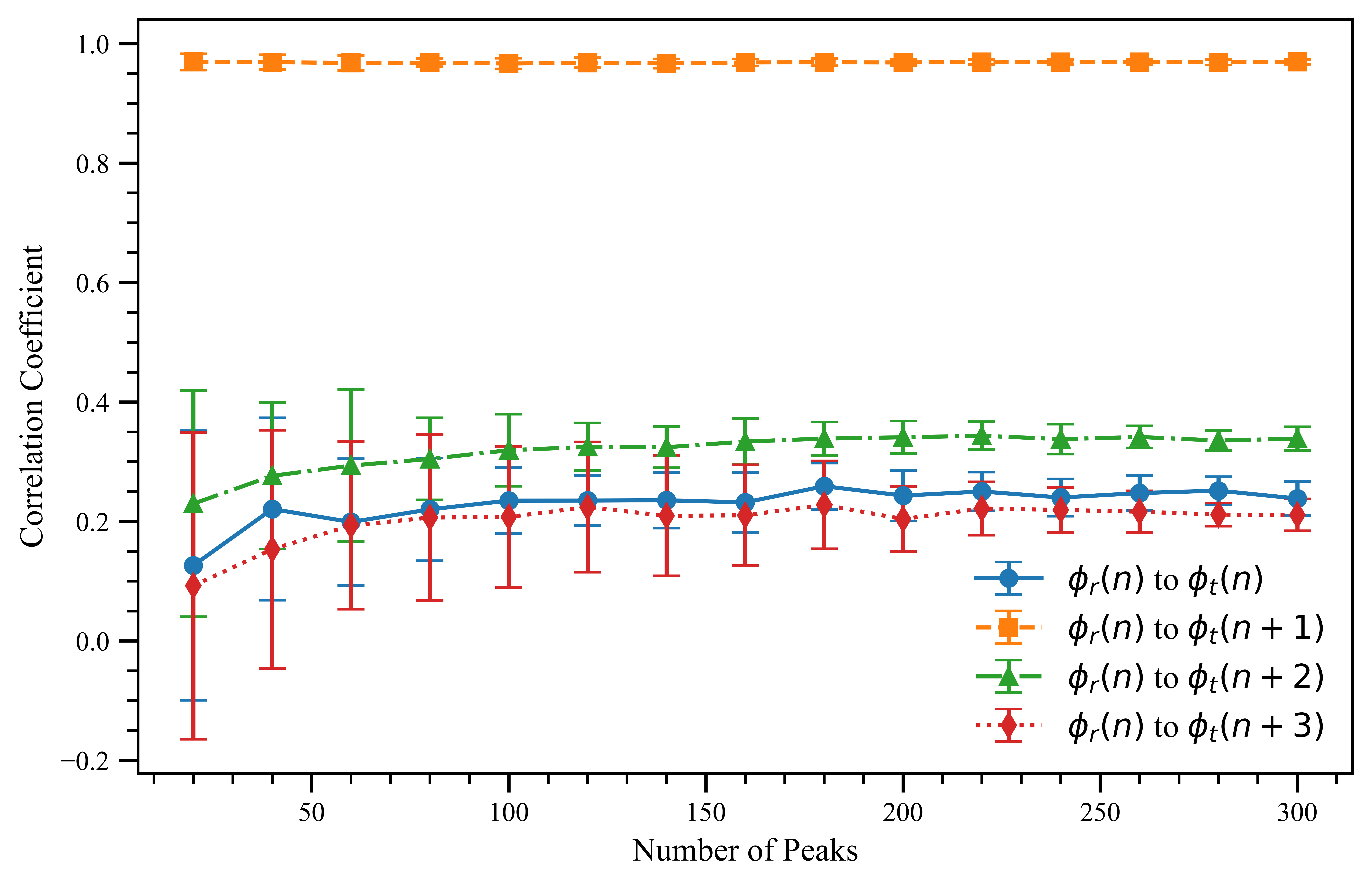}
    \caption{Pearson Correlation coefficients between the polar flux at cycle $n$ with the toroidal flux at cycle $n+1$ (blue/filled-circles), $n+2$ (red/squares), and $n+3$ (black/triangles) computed for a different number of cycles {{from Model II at $\hat{\alpha_0}$ = 2.}} The points show the average values computed from 20 data sets, and the error bar shows the standard deviation.
     }
    \label{fig:corr}
\end{figure*}

{{
We want to add that the observed correlation, as reported by \citet{Muno13}, is based on a limited number of data (10 cycles or 20 data points by separating hemispheres).  And we can observe from \Fig{fig:corr}, that the $n+2$ and $n+3$ (weak) correlations are highly fluctuating when the data point is below about 100.  Hence, the observed no correlation beyond $n+1$ does not utterly confirm that the polar field of the Sun does not have memory beyond one cycle and that the Sun's super-criticality is precisely two times.  Nevertheless, the available data does not contradict the conclusion that is drawn in the previous section about the Sun's supercriticality.
}}

%\red{\textbf{We show that in their model (Model I of \citet{kumar21b}) at $\hat{\alpha}_0=2$, }}which indeed produces a weak multi-cycle correlation beyond $n+2$ cycle, may not display this `weak correlation value' when the number of data points is restricted to 20. As shown in \Fig{fig:corr}, at smaller data points, the correlation value for the polar field of cycle $n$ with the toroidal field of cycle $n+2$ (or of cycle $n+3$) is statistically insignificant; its value varies quite a lot from a negligible to a large one. (If we are lucky enough, we may get a high correlation value.)  Only when the number of data points is more than about 100, we can capture these weak correlations which exist beyond $n+1$ or $n+2$ cycle. \red{\textbf{Therefore, the weak correlation between n cycle polar field (faculae count) and n+2 and n+3 cycle toroidal field (sunspot area) reported in the \citet{Muno13} may be the result of limited data points (10 cycles or 20 data points by separating hemispheres).}} We believe if the sun really has a weak memory of the polar field beyond one cycle, then to establish it, we need good quality data for a much longer duration than the existing duration for the polar faculae count. 

\subsubsection{Gnevyshev–Ohl/Even–Odd Rule}

\begin{figure}
    \centering
    \includegraphics[width=\linewidth]{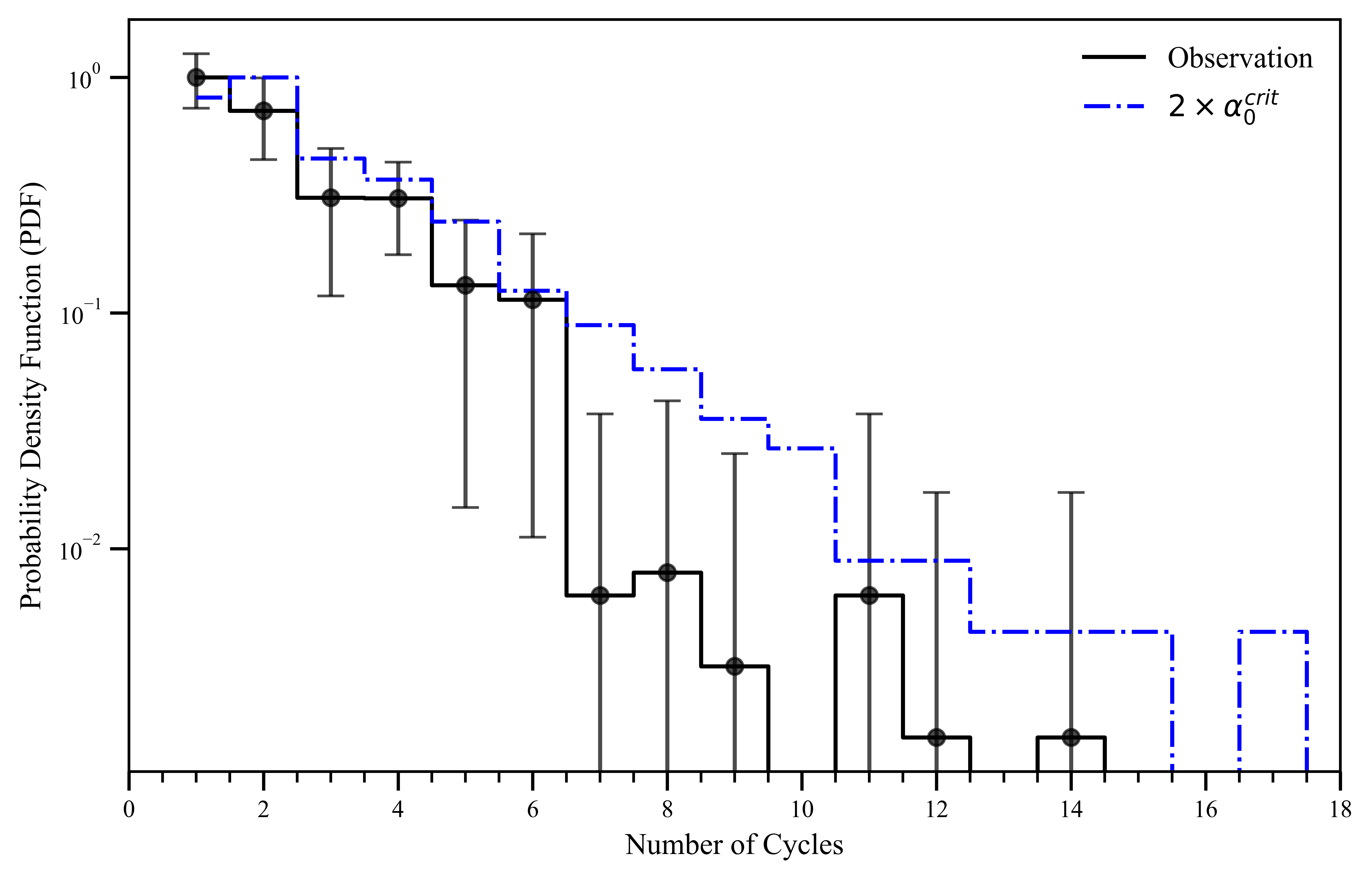}
    \caption{PDF of the duration of the even-odd episodes measured at different values of 
    %dynamo numbers. 
    $\hat{\alpha_0}$ from {{Model IV}} (time delay dynamo model).
    The observation PDF is computed using the last 10 cycles sunspot data and the reconstructed sunspot number of \citet{Usoskin21} with error from \citet{Usoskin22}, excluding the grand minima episodes.}
    \label{fig:enter-label}
\end{figure}

Finally, we shall check another feature of the solar cycle at two times critical dynamo models, and that is Gnevyshev-Ohl, also called the even-odd rule \citep{Hat15, Karak23}. This rule says that if the cycles are arranged in pairs with the even cycle and the following odd cycle, then the sum of the sunspot number in the odd cycle would be higher than the even one. Out of the last 24 solar cycles, this law did not apply for 4-5 and 22–23 pairs \citep{Hat15}.  We check whether our dynamo models at two times supercritical demonstrate this feature. To do so, we compute the probability distribution function (PDF) of the length of the even-odd pair, which is shown by the blue line in \Fig{fig:enter-label}. As for this analysis, we need longer data; we preferred to demonstrate this for the time delay model (Model IV), which is computationally cheap.  To compare with the observed data, we have computed PDF from the observed
{{sunspot number of the last ten cycles in conjunction with the reconstructed sunspot number of the last millennium from \citet{Usoskin21}, which is shown by the solid black line in \Fig{fig:enter-label}. We note that excluding the grand minima phases, we got data points of only seventy-two cycles during normal phases.  Using the errors in the reconstructed sunspot number as reported in \citet{Usoskin22}, we generated $1 \sigma$ errors in the PDF. 
}}
Although the PDF computed from the observed data is not robust due to the limited number of data and a huge uncertainty in the reconstruction \citep{Uso23}, we can say that the model PDF at two times supercritical is not in contradiction to the observed data. 
%We note that with the increase of supercriticality, the probability of holding the even-odd rule is becoming favorable (see dark magenta/dotted line). The reason for this is due to the increasing effect of dynamo growth and nonlinearity. When a cycle becomes weak due to $\alpha$ fluctuations, then immediately, the next cycle becomes strong due to a high dynamo growth rate. The following cycle will be weak because of the strong nonlinearity.

%%%%%%%%%%%%%%%%%%%%%%%%%%%%%%%%%%%%%%%%%%%%%%%%%%
\section{Conclusions}

With the increase of age, the dynamo activity of stars decreases due to the weakening of the rotation rate and, thus, the magnetic field generation efficiency of the dynamo. At some point, stars are expected to reach the subcritical regimes when the dynamo will cease to generate a large-scale magnetic field. We tried to explore how far our Sun is from the critical dynamo transition, in other words, how supercritical our solar dynamo is. To answer this question, we have executed extensive dynamo simulations at different values of the $\alpha$ coefficient as measured by $\hat{\alpha}_0 = \alpha_0 / \alpha_0^{\rm crit}$. By comparing various features of magnetic cycles, namely, the statistics of grand minima/maxima and the recovery rate of Maunder minimum,
%and modeling the amplitudes of the past five solar cycles, 
we conclude that the supercriticality of the solar dynamo is only about two times the critical dynamo ($\hat{\alpha}_0 = 2$). We also showed that this two-times critical dynamo does not contradict the observed correlations between the polar field vs the amplitude of the following cycles as presented in \citet{Muno13} and the probability of the even-odd paring rule. While we have presented our conclusion based on several dynamo models, all are kinematic in nature, and the nonlinearity is specified. However, our conclusion is in line with previous independent studies, which also hint at the weakly supercriticality of the solar dynamo \citep{KN17, CS17, CS19, Ghosh24}.

\section{Acknowledgements} 
The authors gratefully acknowledge the constructive comments and suggestions from the anonymous referee. The authors also thank Vindya Vashishth and Dipanwita Mishra for their valuable comments and discussions. The authors acknowledge the Science and Engineering Research Board (SERB) for providing financial support through the MATRIC program (file no. MTR/2023/000670). 

\section{Data Availability}

We have used the SSN data available at SILSO (www.sidc.be/SILSO/datafiles). Polar field data is taken from Wilcox Solar Observatory (WSO) (http://wso.stanford.edu/Polar.html).

\section{Appendix}
\begin{figure}
    \centering
    \includegraphics[width=\linewidth]{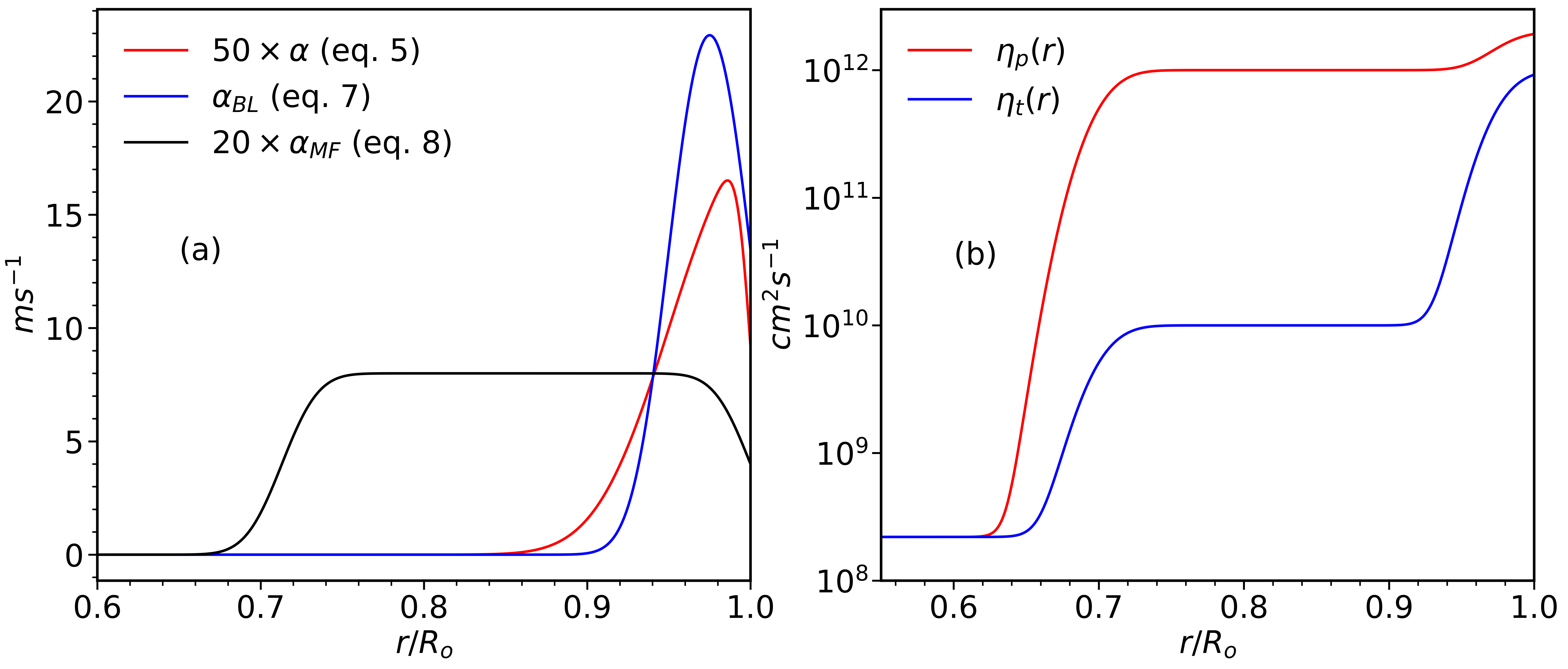}
    \caption{
    {{
    (a): Red and blue curves show the $\alpha$ profile for non-local (for Models I and II), and local (Model III) prescription of \bl\ process. The black curve represents the mean-field $\alpha$ profile for Model III, as discussed in Section 2.1. (b)  Variations of $\eta_p$ (red) and $\eta_t$ (blue) used in Model III.}}}
    \label{fig:profile}
\end{figure}

\begin{figure}
    \centering
    \includegraphics[width=0.65\linewidth]{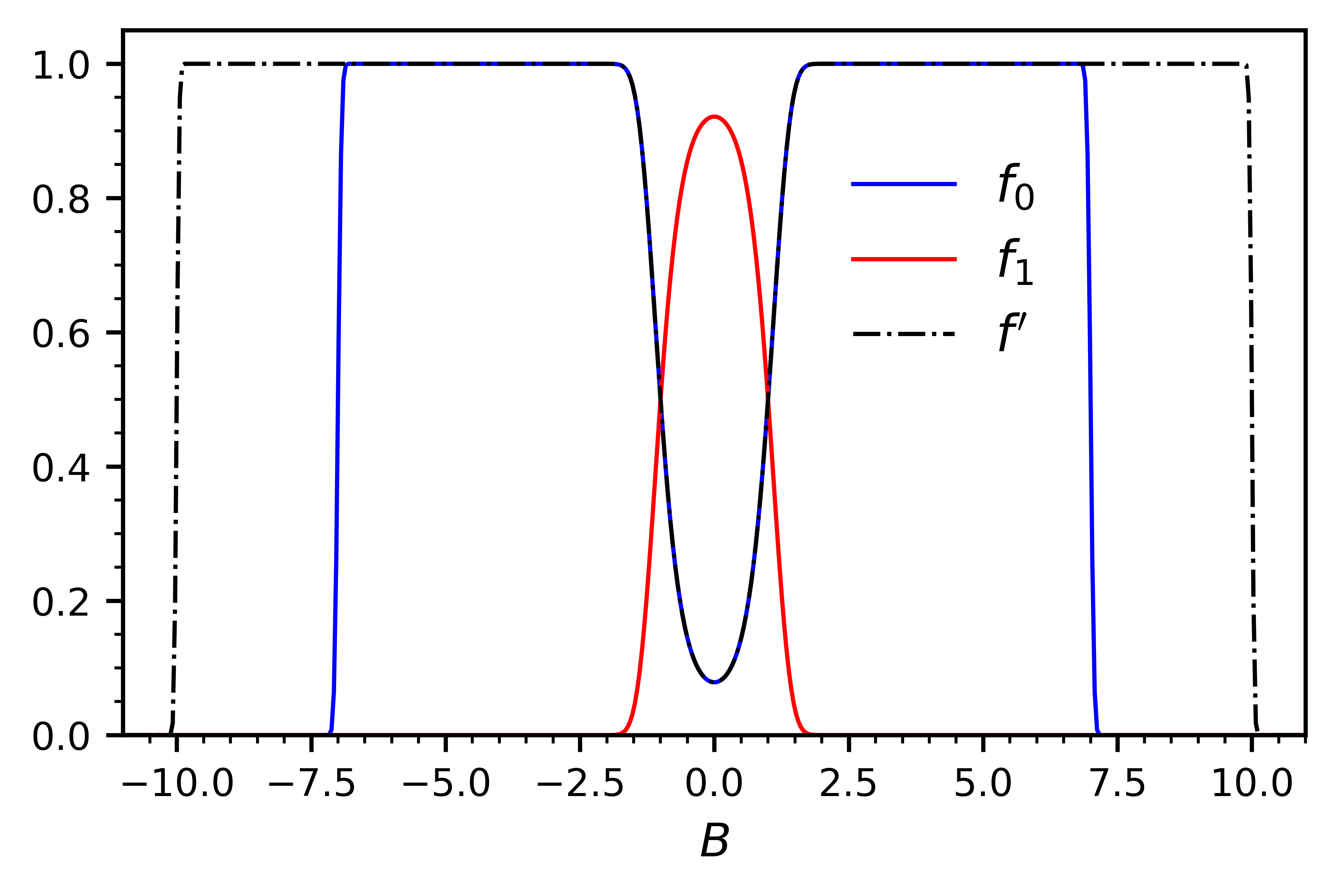}
    \caption{
    {{
    Profiles of the $\alpha$ quenching function used in Models IV and V. Functions $f_0$ and $f_1$ correspond to \bl\ $\alpha$ (blue) and mean field $\alpha$ (red) used in Model IV. The quenching $f'$ corresponds to \bl\ $\alpha$ in Model V is shown in black.}
    }}
    \label{fig:quench}
\end{figure}

{{In this study, we have used different models which use different profiles of $\alpha$, $\eta$, and the quenching. We have already discussed these parameters in the model section in detail. Here, we demonstrate the profiles of different dynamo parameters using \Figs{fig:profile}{fig:quench}.}}
%We note that in Model~III, different values of diffusivity for the poloidal ($\eta_p$) and the toroidal field ($\eta_t$) are used. The justification for this is that the strong toroidal field remains confined in localized regions and experiences more quenching compared to the poloidal one, which remains spread \citep[see][where this idea was introduced]{CNC04}.}} %Additionally, \Fig{fig:quench} shows the variation of the quenching profile in delay dynamo models IV and V.

%%% BIBLIOGRAPHY %%%%%%%%%%%%%%%%%%%%%%%%%%%%%%%%%%%%%%%%%%%%%%%%%%%%%%%%%%%

\bibliographystyle{spr-mp-sola}
     % name your Bibtex file containing your references (.bib)
\bibliography{sola_example_6}

\end{document}